# Superior-catalytic performance of Ni-Co Layered double hydroxide nanosheets for the reduction of p-nitrophenol

Sakshi Kansal[1], Paulomi Singh[1], Sudipta Biswas[2], Ananya Chowdhury[2], Debabrata Mandal[3], Surbhi Priya[1], Trilok Singh[1], Amreesh Chandra[1,2,3*]

[1]School of Energy Science and Engineering, [2]Department of Physics, [3]School of Nano Science and Technology

Indian Institute of Technology Kharagpur, Kharagpur -721302, India

Email: *achandra@phy.iitkgp.ac.in

**ABSTRACT**

Layered double-hydroxides (LDHs) are superior to the conventional (Transition Metal Oxides) TMOs as their lamellar morphology accommodates higher active sites contributing to the facile electron transfer towards the material degradation relations, thus attracts an immense attention for catalysis application. Moreover, the high catalytic activity of LDHs is linked to their facile anion exchange, specific electronic structures, and versatile chemical compositions. Recently, LDHs with bimetallic or ternary combinations are preferably used for hydrogenation of p-nitrophenol, a toxic and carcinogenic pollutant commonly found in industrial wastewater. Herein, a cost-effective co-precipitation fabrication protocol to obtain Ni-Co layered double hydroxide is proposed. The well-defined arrangement of active sites in the 2D structure of LDH, and the synergistic catalytic effect of Ni and Co greatly improves the efficiency for the conversion of p-nitrophenol (p-NP) to p-aminophenol (p-AMP). The catalytic performance at higher temperatures is also shown. The obtained results are further extended to explore the advantage of Ni-Co LDH as an electrocatalyst for Hydrogen Evolution Reaction (HER).

**Keywords:** Layered double-hydroxide; Catalyst; Hydrogen Evolution Reaction (HER)

# 1. Introduction

Layered double hydroxides (LDHs), also known as brucite-like anionic clays with general formula $[M^{II}_{1-x}M^{III}_{x}(OH)_2]_x^+(A^{n-})_{x/n} \cdot yH_2O]$ (where $M^{II}$, $M^{III}$ are divalent, and trivalent metals ions, with $A^{n-}$ as the interlayer anion), are characterized by 2D lamellar within in this structure di- and trivalent cations such as $M^{+2} = Co^{+2}$, $Cu^{+2}$, $Ni^{+2}$, $Mg^{+2}$, and $M^{+3} = Al^{+3}$, $V^{+3}$, $Fe^{+3}$, etc., are orderly distributed in the octahedral voids of edge-sharing hydroxyl layers.[1-2] High redox activities, biocompatibility, prudent synthesis, biodegradable nature, facile substitution of metals in the host layer, and easy anion exchange mechanisms are some of the peculiarities that make them useful for energy devices and catalysis, where electrochemical behaviour decides the applicability of a material.[3-6] The performance can be further enhanced by tuning the 2D-lamellar morphology of LDH, so that it can accommodate more number of charges and facilitates improved redox activities. The high catalytic activity of LDH nanosheets, as compared to metal-oxides counterparts, is linked with their specific electronic structures, versatile chemical compositions, and synergistic contribution from multi-electrocatalytically active metals.[7-8]

For many decades, catalysts based on noble metals are routinely used for the hydrogenation of p-nitrophenol. But their high cost is the major limiting factor. To make environmental remediation more economical, transition metal (especially Ni, Co, Cu) oxides-hydroxides based catalysts have been suggested.[9-11] However, their electroactivity are limited in comparison to noble-metal. Hence, to improve the overall catalytic behaviour, other aspects such as increasing surface area, enhancing catalytically active sites etc. need to be considered. LDHs are used as heterogeneous catalyst in several reaction such as water-splitting, ethanol transesterification, hydrogenation reaction, hydrocarbon oxidation reaction, etc. Among them, LDHs are recently explored for the of p-nitrophenol (p-NP) which is a common toxic pollutant in industrial waste water, to p-aminophenol (p-AMP),that is an important precursor in many pharmaceutical products.[12]

In this paper, the catalytic hydrogenation of p-nitrophenol to p-aminophenol using a Ni-Co based LDH nanosheets is reported. Adopting a simple cost-effective co-precipitation protocol, Ni-Co LDH nanosheets could be synthesized successfully. Further, the structure, morphology, and catalytic activity, of the Ni-Co LDH nanosheets was systematically evaluated. Additionally, kinetic and thermodynamic analysis of the p-nitrophenol conversion reaction were also explored, which helps to define an overall catalytic mechanism for the reaction using Langmuir-Hinshelwood model. The Ni-Co LDH show excellent catalytic property with apparent rate constant ($k_{app}$) = 0.31 min$^{-1}$ at 25º C to $k_{app}$ = 0.65 min$^{-1}$ at 55º C and upto 97% retention of $k_{app}$ after 5 successive cycles at room temperature.

One of the way to produce $H_2$ is water-splitting reaction, which is either catalysed by electrocatalyst or photocatalyst. Herein, bimetallic synergistic effect of two electrocatalytically active transition metals in Ni-Co LDH nanosheets enhances the charge transfer between heterogeneous atoms, and promotes faster water dissociation.[13] This leads to appreciable improvement in the rate of HER. The complete mechanism is also presented in this paper.

## 2. Experimental

### 2.1 Materials used

Cobalt(II) nitrate hexahydrate ($CoNO_3.6H_2O$), nickel(II) nitrate hexahydrate ($NiNO_3.6H_2O$), ammonium hydroxide (28-30%), N-methyl-2-pyrrolidine (NMP), sodium borohydride($NaBH_4$) (95%) were purchased from Merck Specialities Pvt. Ltd. (India), p-Nitrophenol (98%) (p-NP) was acquired from LobaChemie Pvt. Ltd. (India). All the precursors were utilized without further purification.

### 2.2 Synthesis of Ni-Co LDH nanosheets

Ni-Co LDH nanosheets were fabricated using a facile co-precipitation method. 2.5 mmol of each cobalt(II) nitrate hexahydrate (CoNO$_3$.6H$_2$O), and nickel(II) nitrate hexahydrate (NiNO$_3$.6H$_2$O), were taken in 50 ml DI water, and the concentration of the solution was maintained to 0.1 mol L$^{-1}$. Further, the solution was stirred (1100 rpm) at room temperature and ammonia, as a precipitating agent, was added gradually to the solution drop wise for achieving a pH of 9. This solution was further kept for stirring for 12 h, followed by filtration, and washing by distilled water and ethanol several times. Subsequently, the synthesized material was dried at 60° C for 12 h, to obtain the final desired product. The involved chemical reaction are given below: [14-15]

$$[Ni(NO_3)_6].6H_2O + 6NH_3 \leftrightarrow [Ni(NH_3)_6]^{+2} \quad (1)$$

$$[Co(NO_3)_6].6H_2O + 6NH_3 \leftrightarrow [Co(NH_3)_6]^{+2} \quad (2)$$

$$NH_3 + H_2O \leftrightarrow NH_4^+ + OH^- \quad (3)$$

$$(1-x)Ni[(NH_3)_6]^{+2} + xCo[(NH_3)_6]^{+2} + 2OH^- \leftrightarrow Ni_{1-x}Co_x(OH)_2 \quad (4)$$

$$Ni_{1-x}Co_x(OH)_2 + xNO_3^- + yH_2O \leftrightarrow Ni_{1-x}Co_x(OH)_2(NO_3^-)_x \cdot yH_2O + xe^- \quad (5)$$

## 2.3 Materials characterization

The phase formation of the samples was affirmed by the investigation of X-ray diffraction data collected using a Rigaku X-ray diffractometer with CuK$_\alpha$ incident radiation (λ= 0.15406 nm), in the 2θ range 5-80°. Scanning electron (SEM CARLZEISS SUPRA 40, 20 kV), transmission electron microscopy (TEM FEI-TECNAI G220S-Twin operated at 200 kV) and high resolution TEM (JEOL JEM-2100) were utilized for investigating the particle size, shape, and morphology. X-ray photoelectron spectroscopy (XPS) measurements data was acquired utilizing PHI-5000 VERSAProbe II X-ray photoelectron spectrometer that had monochromatic Al Kα as the incident photon energy. The Brunauer–Emmett–Teller (BET) surface area and porosity were measured with a Quantachrome Autosorb-iQ/MP-XR surface area analyzer. The specific surface area (S$_{BET}$), and the pore size distribution were evaluated by using Brunauer-Emmett-Teller (BET) method upon the adsorption curve, and the Barrett-Joyner-Halenda (BJH) method, respectively. UV-Vis spectrophotometer (Avantes Starline AvaSpec-ULS3648) was used to record the UV-Vis absorption spectra and FTIR performed using SHIMADZU IRSiprit in the range of 200-600 nm.

### *2.4. Catalytic measurement*

To investigate the performance of Ni-Co LDH nanosheets as a catalytic agent, initially, 60 mg of $NaBH_4$ ($8 \times 10^{-2}$ M), and 20 mL aqueous solution of p-nitrophenol ($10^{-4}$ M) were mixed. The absorption maximum of p-nitrophenolate ions at 400 nm was recorded. Consequently, varying concentrations of Ni-Co LDH catalyst i.e. 0.25, 0.5, to 1 mg mL$^{-1}$ was added to the above mixture, and time-dependent UV-Vis spectra was recorded in the range of 200-500 nm. Further, keeping the concentration of Ni-Co LDH nanosheets same, the rate of reaction for conversion of p-nitrophenol to p-aminophenol was recorded at temperatures ranging from 25º C to 55º C. Additionally, the reusability and stability of the catalyst were analysed taking separated catalyst from one cycle and to reuse it again for the catalytic reduction of p-nitrophenol upto 5 cycles.

### *2.5 Electrochemical measurement*

The electrochemical measurements such as linear sweep voltammetry (LSV), chronoamperometry, and impedance spectroscopy (EIS), and were accomplished utilizing the Metrohm Autolab (PGSTAT302N) using a standard three-electrode cell configuration in 1 M KOH aqueous electrolyte taking platinum electrode and Ag/AgCl/3.0 M KCl as the counter and reference electrodes, respectively. For the fabrication of Ni-Co LDH electrodes 80 wt% of Ni-Co LDH, 10 wt% of activated carbon (AC), and 10 wt% polyvinylidenefluoride(PVDF) were mixed using acetone as a mixing media and the mixture was kept for stirring at 60º C for 6 h to obtain a homogeneous slurry. The as-synthesized slurry was dropcasted onto Ni-foam (1cm x 1cm). The active mass of the working electrode was ~1mg cm$^{-2}$. Furthermore, the electrocatalytic HER activity of the material was measured in 1 M KOH solution at room temperature. In all measurements, Ag/AgCl, reference electrode was calibrated relative to the reversible hydrogen electrode (RHE), using the formula $E(RHE) = E(Ag/AgCl) + 0.197 V + 0.059 \times pH$. Linear sweep voltammogram (LSV) were obtained at a scan rate of 5 mV s$^{-1}$. The long-term stability of Ni-Co LDH catalyst were investigated by chronopotentiometric measurements. The EIS measurements were performed in the frequency range from 0.01 Hz to 100 kHz.

## 3. Results and discussion

### *3.1 Physiochemical characterization*

Figure 1(a)depicts the XRD-diffraction pattern for Ni-Co LDH nanosheets, which comprised of four peculiar diffraction peaks appearing at 2θ ~ 11.3º, 22.5º, 34.2º, and 60.2º. These could be designated to (001), (002), (100) and (110), respectively, reflections of the hydrotalcite-like structure, using the standard JCPDS card no.-33-0429.[16] No impurity peaks due to α-Ni(OH)$_2$ or α-Co(OH)$_2$ were discernible, which clearly indicated formation of single phase material. The basal plane spacing d$_{(001)}$ for as-prepared Ni-Co LDH was ~ 0.79 nm, which was more than the interlayer spacing of pure α-Ni (OH)$_2$ (0.756 nm) and α-Co (OH)$_2$ (0.776 nm).[17]

In Figure 1(b), a typical SEM micrograph of as obtained Ni-Co LDH is illustrated, which highlights the morphology of Ni-Co LDH as nanosheets. The SEM micrographs also indicated that the Ni-Co LDH nanosheets also had highly porous surface. Furthermore, figure S1 depicts the energy dispersive X-ray (EDX) analysis of as-fabricated Ni-Co LDH nanosheets. The study corroborated that, the ratio of Ni: Co was 1:1, which was similar to the feeding ratio of the precursor salts which demonstrate co-precipitation as an efficient method for the preparation of Ni-Co LDH nanosheets. TEM micrograph of Ni-Co LDH nanosheets, shown in figure S2 (a) reconfirmed the layered structure while figure S2 (b), showing the selected area diffraction pattern (SAED) pattern, indicated the amorphous nature of the material.

It is essential to analysis the specific surface area of a catalyst as it can directly impacts the catalytic efficiency. For determining the specific surface area of Ni-Co LDH nanosheets, $N_2$ adsorption/desorption analysis was conducted and the corresponding result is shown in Figure 1(c). The curve was similar to type-IV isotherm, with a hysteresis loop ($p/p_0 > 0.4$) which indicated towards mesoporous nature of the sample.[18] The specific surface area of LDH was 123 $m^2 g^{-1}$ and from pore-size distribution curve (inset), with pore diameter 3.71 nm. A sharp single peak obtained in pore size distribution curve indicated towards monodispersed pores on the layered surface. In Figure 1(d), FTIR spectrum is highlighted. The peak at 3443 $cm^{-1}$ indicated towards the presence of O-H. Additionally, peak at 1648 $cm^{-1}$ for the bending mode of water molecules in the interlayer, and $NO_3^-$ asymmetric stretching peak at 1381 $cm^{-1}$ were also clearly visible. Peaks beyond 800 $cm^{-1}$ corresponds to the bending vibrations of metal-oxygen (M–O) bond, confirming the presence of Ni-Co LDH.[19] Figure(e), and (f) represents the X-ray photoelectron spectroscopy (XPS) spectrum of Ni-Co LDH. The XPS spectrum showed the presence of Ni 2p and Co 2p peaks. The Ni 2p spectrum shown in Figure 1 (e) had two significant peaks at 855.5 and 873.2 eV, corresponding to Ni $2p_{3/2}$ and Ni $2p_{1/2}$, respectively. The satellite peaks at 861.3 and 879.1 eV could be attributed to the presence of signal $Ni^{2+}$ ion. The $Co_{2p}$ spectrum in Figure 1(f) comprised of two shakeup satellites, and two spin-orbit doublets. One pair of binding energies, centred at 780.2 and 795.4 eV, attributes to $Co^{3+}$. The other pair, at the higher energy values of 782.3 and 797.3 eV, could be designated with to $Co^{2+}$, in addition to the linked two satellite peaks at 785.8 and 802.9eV, respectively. These outcomes demonstrate that there exist two oxidation states for cobalt (i.e., $Co^{2+}$ and $Co^{3+}$) in the composite.[20-21] This can synergistically contributed in the catalytic behaviour.

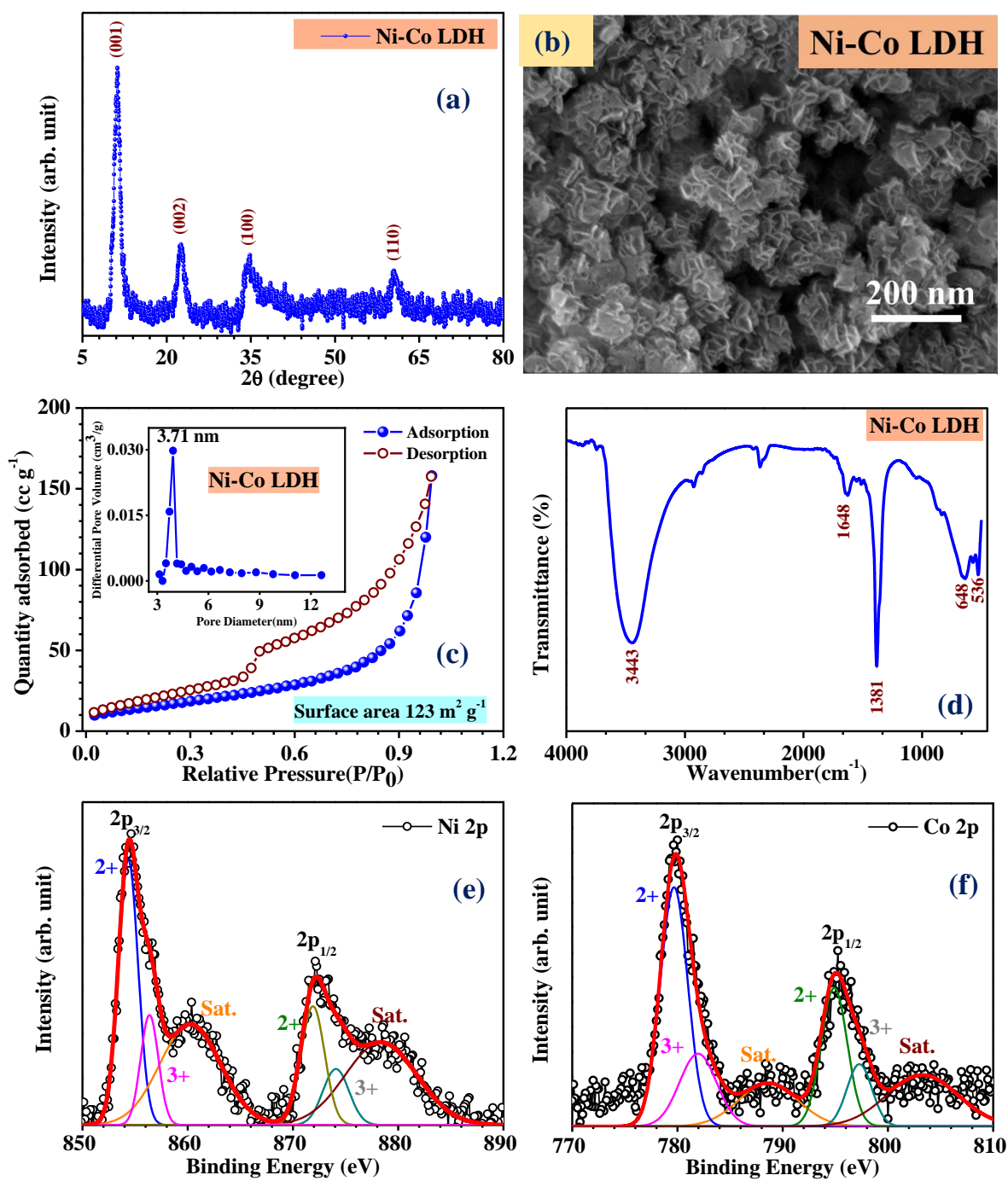

Figure 1 (a) XRD pattern, (b) SEM image, (c) N$_2$ adsorption desorption curve (d) FTIR curve (e, f) XPS of Ni, and Co, respectively.

**3.2 Catalytic activity of Ni-Co LDH nanosheets for p-nitrophenol reduction**

The catalytic ability of as-prepared Ni-Co LDH nanosheets were explored for the hydrogenation of the p-nitrophenol (p-NP) using excess NaBH$_4$ as a reducing agent. This reduction process was examined by analysing the absorption spectra evolution of UV-Vis as a function of time. As soon as NaBH$_4$ was added to the p-NP solution, the colour of solution turned to intense yellow due to the change in pH from acidic to highly basic. The significant peak of p-NP at 317 nm indicated a bathochromic shift to 400 nm, indicating towards the formation of p-nitrophenolate anions as an intermediate. This can be clearly seen in Figure 2(a).

The hydrogenation reaction of p-nitrophenol to p-aminophenol(($E°_{(p-NP/p-AMP)}$= - 0.76 V) by NaBH$_4$ ($E°_{(H3BO3/BH4-)}$ = -1.33 V) is thermodynamically achievable but it is a kinetically sluggish due to electrostatic repulsion between the intermediate anion (p-nitrophenolate anion) and BH$_4^-$.[22] This can be inferred by analysing Figure 2(b), as the peak at 400 nm remained unaltered even after 3 h. Hence, an efficient catalyst is required to speed the conversion reaction. In this work, 60 mg of NaBH$_4$ was added to 20 mL of p-NP solution ($10^{-4}$ M). Subsequently, varying amounts of catalyst was added to this solution and absorbance change, at 400 nm, was carefully monitored. This is shown in Figure 2(c, d) and Figure 3(a). The total time required to complete the reduction was also recorded, which is indicated by a hypochromic shift of peak at 400 nm to 300 nm for p-aminophenol (p-AMP) in absorption spectra and also the decolourization of the yellow solution. The apparent rate constants were evaluated for the reduction reaction of p-NP at various concentration of catalyst, the Beer-Lambert Law was used, which is given by:

$$A = \varepsilon C l \qquad (6)$$

where, A is absorbance at particular wavelength, $\varepsilon$ = molar absorption coefficient, C is the concentration of the solution, $l$ and is the length of cuvette. Since, the nature of the solution and the length of cuvette was same in all experiments, absorbance would be directly proportional to the concentration and ratio of $C_t/C_o$ can be written as equal to the ratio of $A_t/A_o$, where $C_o$, $C_t$ are the concentrations of p-NP, and $A_o$, $A_t$ are the absorbance at 400 nm for p-nitrophenol solution at time t = 0 s and at any other time interval, respectively.[23] The change of $C_t/C_o$ with time (t) is shown in Figure 2(e). Further, as the concentration of NaBH$_4$ was higher than the p-nitrophenol, the overall rate of reaction will follow pseudo-first order kinetics and can be written as:[24-25]

$$-\frac{dC}{dt} = k_{app}t \qquad (7)$$

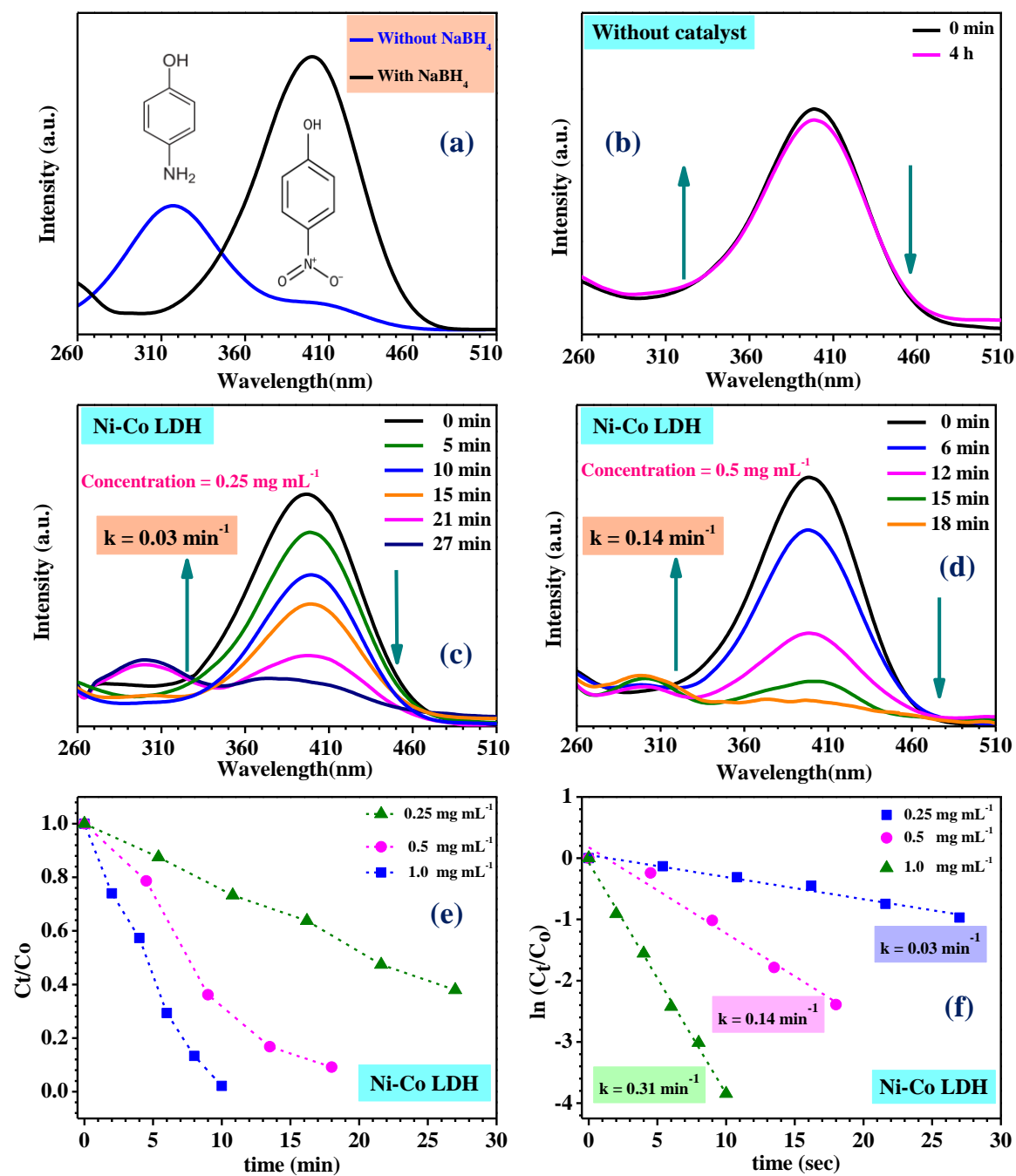

Figure 2 Absorption spectra (a) with and without catalyst (b) without catalyst (c) with catalyst with conc. 0.25 mg mL$^{-1}$, (d) with catalyst with conc. 0. 5 mg mL$^{-1}$ (e) $C_t/C_o$ vs t, and (f) Pseudo first order kinetics

$$\ln\left(\frac{C_t}{C_o}\right) = -k_{app}t \qquad (8)$$

where, t is the reaction time, and $k_{app}$ is the apparent rate constant. A good linear relationship between $\ln(C_t/C_o)$ and t can be seen in Figure 2(f). Therefore, $k_{app}$ would be calculated from above mentioned equation. Table S1 list the different time intervals needed to complete the reduction of p-nitrophenol, at different concentration of Ni-Co LDH, and the corresponding apparent rate constants. The highest catalytic activity was shown by the Ni-Co LDH nanosheets when the concentration was taken as 1 mg mL$^{-1}$ as the catalysis would be completed in 10 minutes at room temperature, with a rate constant value of 0.31 s$^{-1}$.

### 3.3 Thermodynamic analysis of Ni-Co LDH nanosheets for p-nitrophenol reduction

For analysing the performance of the Ni-Co LDH nanosheets as a catalyst and to calculate the thermodynamic parameters for the p-nitrophenol reduction reaction, the experiment was performed using the 1 mg mL$^{-1}$ of the catalyst, while keeping concentration of p-NP, and NaBH$_4$ same. The experiments were performed at temperatures ranging from 25 to 55º C. Figure 3 (a-d) shows the absorption spectra for p-nitrophenol conversion at varying temperature. Moreover, the $C_t/C_o$ Vs t curve are shown in Figure 4 (b). Further, $\ln(C_t/C_o)$ Vs t was plotted and is shown in Figure 4 (a). Using this, $k_{app}$ values, at different temperature, were evaluated. Table S2 shows, the time required to complete the reaction at various temperature with their respective $k_{app}$ values. It was evident that, as the temperature increased, the reaction become faster and the apparent rate constant also increased.

Further, the catalytic efficiency of Ni-Co LDH was verified and evaluated at different temperature by calculating two parameters i.e. Turn Over Number (TON) and Turn Over Frequency (TOF) using the following equations:[24, 26]

$$TON = \frac{No.\,of\ molecules\ of\ substrate\,(p-NP)}{per\ g\ of\ catalyst} \qquad (9)$$

$$TOF = \frac{TON}{t} = \frac{No. of\ molecules\ of\ substrate(p-NP)}{per\ g\ of\ catalyst * time\ taken} \quad (10)$$

The TOF values of p-nitrophenol reduction by Ni-Co LDH nanosheets are given in Table S2. It was observed that, at room temperature, 1 mg of Ni-Co LDH was able to convert $1.21*10^{20}$ molecules of p-nitrophenol to p-aminophenol in 10 min. Furthermore, to explain the rate kinetics of any reaction, it is mandatory to determine the activation energy. The overall activation energy required for the conversion of p-nitrophenol to p-aminophenol was determined using the *Arrhenius rate law*:

$$lnk_{app} = lnA - \frac{E_a}{RT} \quad (11)$$

where, $k_{app}$ is the apparent rate constant for a particular reaction ($s^{-1}$), A is the pre-exponential factor or frequency factor, which indicated the number of collisions occurring between the reactants molecules irrespective of their possessed kinetic energy ($s^{-1}$), $E_a$ is the activation energy (kJ $mol^{-1}$) and T is the temperature (K). From the slope of the $lnk_{app}$ vs $T^{-1}$ curve, activation energy could be estimated as 20.2 kJ $mol^{-1}$, as shown in Figure 4 (c). Activation energy also reveals the relation between the apparent rate constants with change in temperature. For higher value of activation energy, the rate constant will be more influenced by change in the temperature.[27] A slight change in the rate constant values were observed with increasing temperature.

Furthermore, to compute the thermodynamic parameters, Transition-State theory was applied, which expresses the activation process in terms of thermodynamic functions such as entropy of activation ($\Delta S^{\#}$), activation enthalpy ($\Delta H^{\#}$), and Gibbs energy of activation ($\Delta G^{\#}$). The Eyring equation, derived from Transition-State theory, can be written as:[27]

$$\ln\left(\frac{k_{app}}{T}\right) = -\frac{\Delta H^{\#}}{R} + \ln\left(\frac{k_B}{h}\right) + \left(\frac{\Delta S^{\#}}{R}\right) \quad (12)$$

where, $k_{app}$ is the apparent rate constant (s$^{-1}$), $\Delta H^{\#}$ is the activation enthalpy, $\Delta S^{\#}$ is the entropy of activation, and $k_B$ and $h$ are Boltzmann constant and Planck's constant, respectively. From the slope and intercept of the ($k_{app}$/T) vs T$^{-1}$ curve, depicted in Figure 4 (d), the $\Delta H^{\#}$, and $\Delta S^{\#}$ values were estimated as 17.58 kJ mol$^{-1}$ and -195.7 J mol$^{-1}$, respectively. The positive value of $\Delta H^{\#}$ indicated towards the endothermic nature of the reduction reaction. This corroborates our previous result that, on increasing the temperature, the rate constant increases. Further, the value of $\Delta S^{\#}$ reveals the association of the reactant molecule as adsorbate at the catalyst surface i.e. Ni-Co LDH nanosheets. The negative value of the $\Delta S^{\#}$ suggested that the randomness of the system decreases as the reaction proceeds with the decrement in number of p-nitrophenol molecules.[28] Additionally, $\Delta G^{\#}$, at each temperature, were evaluated using the relation given below and are listed in Table S2.

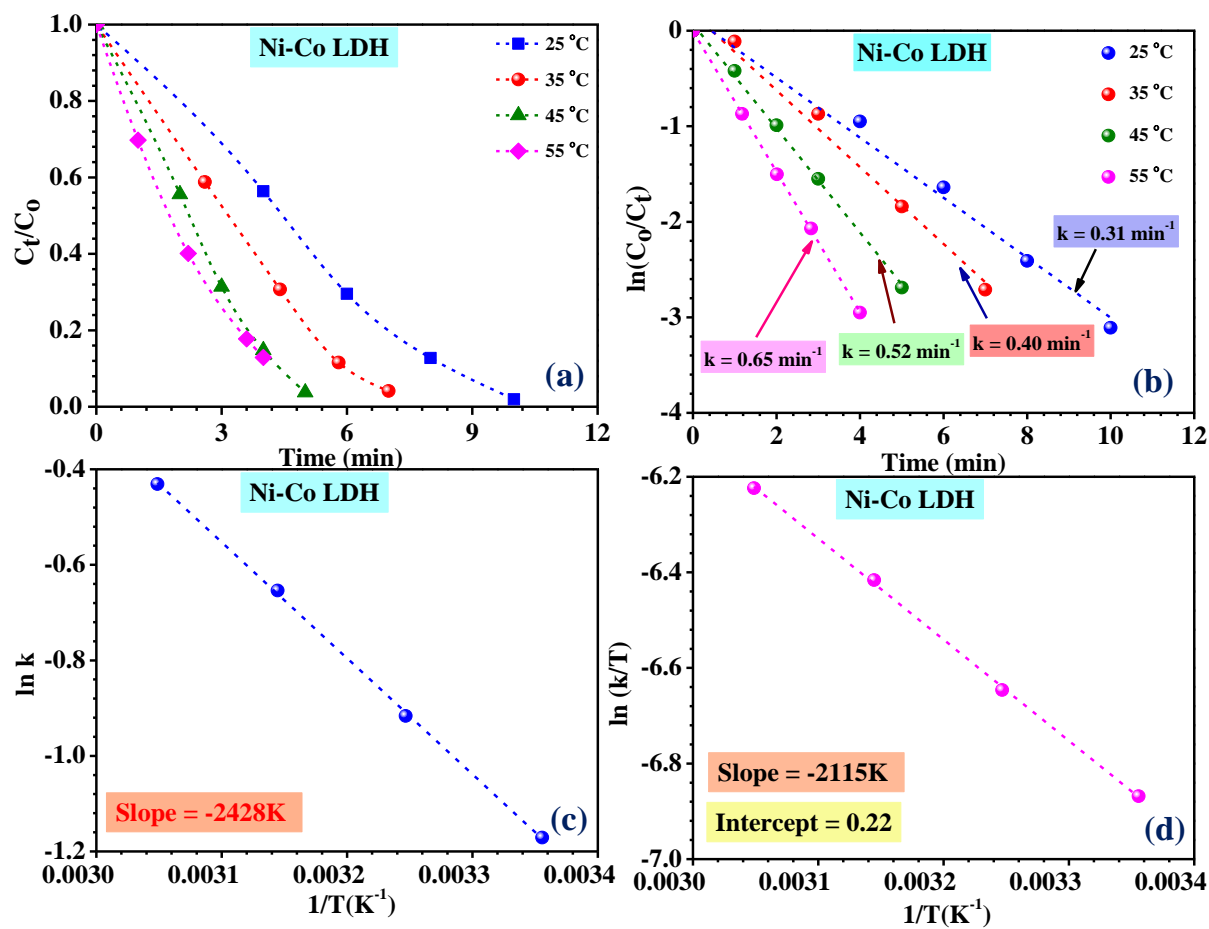

Figure 3 (a) $C_t/C_o$ vs t (b) Pseudo first order kinetics (c), Arrhenius curve (d) Erying plot

$$\Delta G^{\#} = \Delta H^{\#} - T\Delta S^{\#} \qquad (13)$$

From the data, the value of ΔG$^{\#}$ was positive and increased with temperature increases, indicating towards the endergonic nature of the p-nitrophenol conversion reaction.

**3.4 Mechanism of p-nitrophenol reduction by Ni-Co LDH nanosheets as catalyst**

There can be two plausible mechanisms for the conversion reaction of p-nitrophenol to p-aminophenol by heterogonous catalyst:

(i) It could be explained by using Langmuir–Hinshelwood model, where the intermediate (p-nitrophenolate ions) and $BH_4^-$ both are adsorbed on the surface of Ni-Co LDH nanosheets, and then the electron transfer occurs and,[27, 29]

(ii) By Eley–Rideal model[27], where in which only $BH_4^-$ got adsorbed on the surface of LDH, releases the hydride ion that get adsorbed on the surface of LDH and later reacted with p-nitrophenolate ions in the solution.

In order to deduce the plausible mechanism, the dependence of the apparent rate constant ($k_{app}$) on the concentration of p-NP was analysed by keeping the amount of catalyst and NaBH$_4$ same. For Eley–Rideal model, the rate constant should increase with increase in the concentration of p-nitrophenol. If the values decreased with p-nitrophenol concentration then the first model i.e. the Langmuir–Hinshelwood model will justify the overall reduction mechanism by Ni-Co LDH nanosheets.[26, 30] Figure 5 (a) shows an almost inverse relationship between the apparent rate constants value with increase of p-nitrophenol concentration, indicating towards the Langmuir–Hinshelwood mechanism. Also, analysis of the rate constant with increment in the concentration of NaBH$_4$ was performed by keeping the concentration of p-nitrophenol and catalyst same, and is shown in Figure 5 (b). From the curve, it can be observed that, with the increment in NaBH$_4$ concentration, firstly the $k_{app}$ increased, upto a certain value before showing saturation. The decrement in $k_{app}$ values, with increase in p-nitrophenol concentration and non-linear dependence of k$_{app}$ over the concentration of NaBH$_4$, indicated towards the successive competition between both the reactants for the adsorption on the surface of the catalyst.

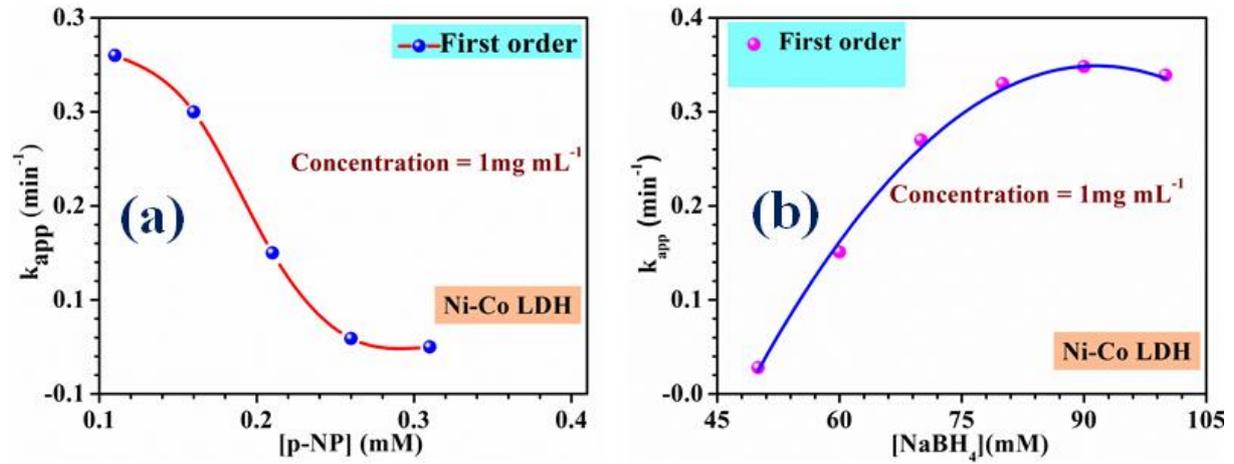

Figure 4 (a), and (b) Effect of variation of p-NP and NaBH$_4$ concentration on the value of apparent rate constant (k$_{app}$)

Following the Langmuir-Hinshelwood model, at first, the borohydride (BH$_4^-$) would adsorb on the surface of Ni-Co LDH nanosheets (Eq 9 ). The subsequent electron transfer to H$_2$O, in order to produce an active hydrogen species, would drive its adsorption onto the surface of catalyst, (Eq 10). Therefore, p-nitrophenolate molecules, adsorbed on the surface of Ni-Co LDH nanosheets (Eq 11) were hydrogenated by the active hydrogen species to produce a stable intermediate "p-hydroxylaminophenol" (Eq 12). Finally, formation of p-aminophenol, with consequent desorption of it from the surface of Ni-Co LDH nanosheets would occur (Eq 13). The whole process can be understood by reactions scheme (Figure 6) mentioned below:

$$BH_4^- + S_{LDH} \leftrightarrow BH_4^-.S_{LDH} \quad (14)$$

$$BH_4^-.S_{LDH} + 2H_2O \rightarrow BO_2^-.S_{LDH} + 8H^\#.S_{LDH} \quad (15)$$

$$p-NP^- + S_{LDH} \leftrightarrow p-NP^-.S_{LDH} \quad (16)$$

$$p-NP^-.S_{LDH} + 6H^\#.S_{LDH} \rightarrow p-AMP^\#.S_{LDH} \quad (17)$$

$$p-AMP^\#.S_{LDH} \rightarrow p-AMP + S_{LDH} \quad (18)$$

where $S_{LDH}$ is catalyst surface, $BH_4^-.S_{LDH}$ is borohydride adsorbed on catalyst surface, $H^\#.S_{LDH}$ is active hydrogen species adsorbed on the surface, p-NP⁻ is p-nitrophenolate ion, p-NP⁻.$S_{LDH}$ is p-nitrophenolate ion adsorbed on catalyst surface, p-AMP$^\#$.$S_{LDH}$ is p-hydroxylaminophenol, p-AMP is p-aminophenol.[24, 26]

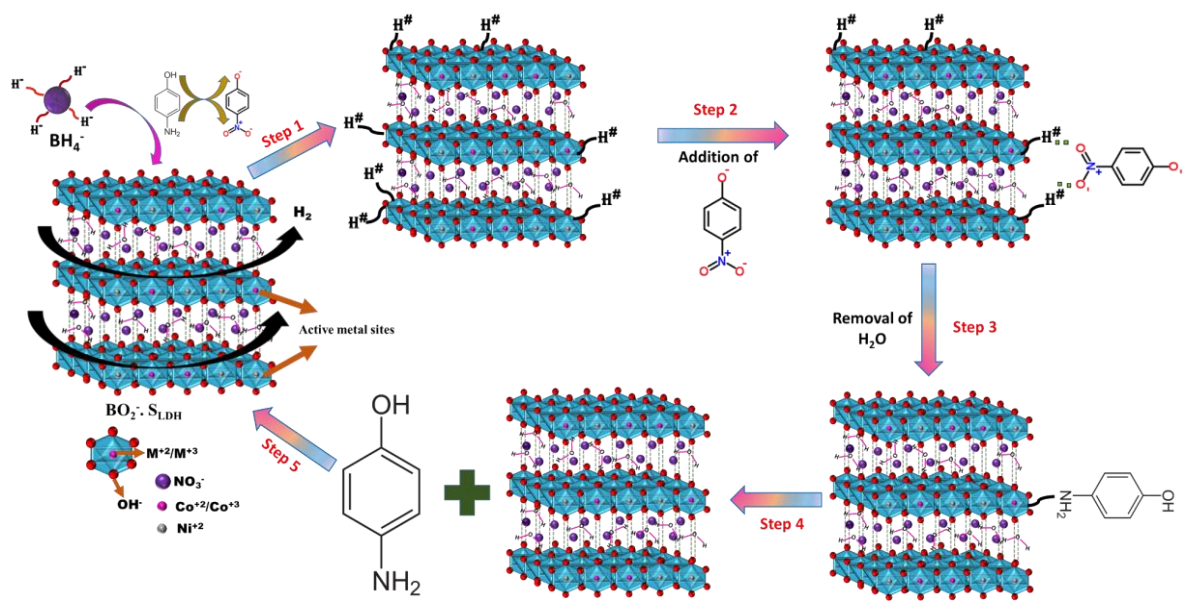

Figure 5 Mechanism of p-nitrophenol reduction by NaBH$_4$ as reducing agent and Ni-Co LDH as catalyst

## 3.5 Reusability of Ni-Co LDH catalyst for p-nitrophenol reduction reaction:

The reusability is one of the most notable factor to analyse the performance of catalyst in practical applications. Therefore, after completion of the first set of catalytic reaction, the Ni-Co LDH nanocatalyst was separated and filtered using centrifuge, washed with water-ethanol and then kept for drying at 60º C for 24 h. Subsequently, the reused nanocatalyst was weighed and added for catalysis in an equivalent amount of p-nitrophenol-NaBH$_4$ solution. The process was repeated for 5 cycles and the result is shown in the Figure S4(a). The Ni-Co LDH nanosheets retained good catalytic activity even after 5 successive cycles with 97% retention of the k$_{app}$ value as compared to first cycle. Additionally, XRD data for the sample after 5 cycles was collected and compared with the initial XRD of Ni-Co LDH. No phase change was observed, indicating towards high structural stability (Figure S4(b)). However, the crystalline size of the nanoparticles increased from 6.24 nm to 7.64 nm, which accounts as major reason for the decrement of the catalytic activity after successive cycling.[31]

**3.6 Catalytic performance of Ni-Co LDH for Hydrogen Evolution Reaction (HER)**

On evaluating the superior catalytic performance of Ni-Co LDH nanosheets, electrocatalytic ability of Ni-Co LDH nanosheets was also investigated for the hydrogen evolution reaction (HER) in alkaline medium. The mechanism of hydrogen evolution reaction (HER) using an electrocatalyst followed *Volmer-Heyrovsky pathway*, which can be explained in two steps:[32-33]

(i) The electrocatalyst activates and alter the O-H bond to produce $H_{ads}$ on its surface (Eq 14)

(ii) $H_{ads}$ desorp and react with water in order to obtain $H_2$ gas (Eq 15).

$$E_{LDH} + H_2O + e^- \rightarrow E_{LDH} - H_{ads} + OH^- \text{ (Volmer step)} \quad (19)$$

$$H_2O + E_{LDH} - H_{ads} + e^- \rightarrow H_2 + E_{LDH} + OH^- \text{(Heyrovsky step)} \quad (20)$$

$$2E_{LDH} - H_{ads} \rightarrow 2E_{LDH} + 2H_2 \text{ (Tafel reaction)} \quad (21)$$

where, $E_{LDH}$ is Ni-Co LDH as electrocatalyst, and $E_{LDH}$-$H_{ads}$ is hydrogen adsorbed on surface of electrocatalyst.[13, 34] As LDHs are very stable in alkaline medium, they can be beneficial as catalyst for the HER. The electrocatalytic ability of Ni-Co LDH nanosheets was evaluated in three-electrode system taking Ag/AgCl as reference, Ni-Co LDH nanosheets@Ni-foam (Ni-Co LDH@NF) as working electrode, and Pt electrode as counter electrode in 1 M KOH. Figure 6 (a) represents the LSV curves for the Ni-Co LDH@NF and bare Ni-foam for HER. It can be clearly observed that Ni-Co LDH@NF display a high catalytic activity, better than the earlier reported metal oxide-hydroxides based-electrocatalysts mentioned in Table S3. The 2D-lamellar structure of LDHs provides high surface area, and also the synergistic effect of two electrocatalytically active metals Ni and Co improves the electron transfer process and increases the rate of HER.[35-36] Furthermore, the overpotential of 247 mV was observed at the current density of 10 mA cm$^{-2}$. Figure 6(b) depicts the Tafel plot fitted by Tafel equation (η =

a+ b log|j|) based on the Ni-Co LDH LSV curve presented in Figure 6(a). The value of Tafel slope was less as compared, to earlier reported articles (Si Table S3). Further, the cyclic stability of Ni-Co LDH nanosheets was also analysed in order to ensure long cycle life. Chronoamperometry was performed at a constant voltage of -0.15 V for >12 h and the results are shown in Figure S5. Only a slight decrease in current density was observed even after 12 h, which highlights an appreciable stability of Ni-Co LDH as electrocatalyst for HER.[37-38] Additionally, stability was ensured by comparing the LSV curves of Ni-Co LDH before and after 1000 CV cycles. This displayed nearly no change in the electroactivity in Figure 6(c). Finally, to evaluate the reaction kinetics at the electrode/electrolyte interface, electrochemical impedance spectroscopy (EIS) was estimated and charge transfer resistance was calculated for Ni-Co LDH@NF and compared it with bare Ni-foam.[39-40] It was evident that more the $R_{ct}$ value

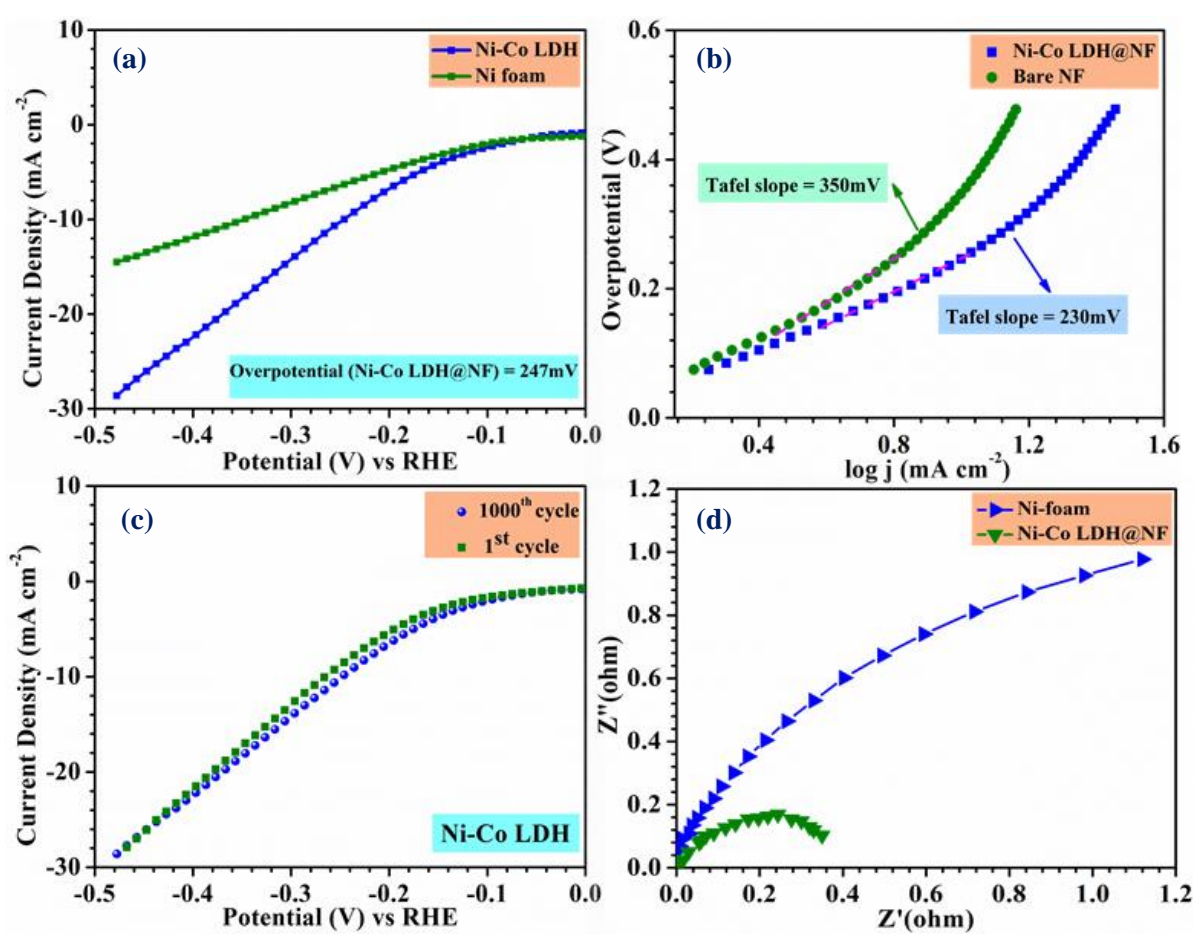

Figure 6 (a) LSV Curve (b) Tafel curve (c) Cyclic variation, and (d) EIS for Ni-Co LDH vs Ni-foam with concentration of (a) p-NP (b) NaBH$_4$ more sluggish will be the rate kinetics of electron transfer at electrode-electrolyte interface as highlighted in Figure 6(d).

## 4. Conclusion

In summary, it is observed that Ni-Co LDH nanosheets can be synthesized via using a facile precipitation method. Owing to the synergistic effect produced by two electrocatalytically active transition metals, the sample show high catalytic performance for the reduction of p-Nitrophenol (p-NP) with good reusability. The best catalytic activity shown by Ni-Co LDH was at concentration of 1 mg mL$^{-1}$, where it could facilitate conversion of p-nitrophenol to p-Aminophenol(p-AMP) in just 10 minutes. The activation energy for the reaction was 20.2kJ mol$^{-1}$, with positive value of Gibbs energy of activation. This proves the necessity of a catalyst for carrying out the hydrogenation of p-NP. The reduction of p-NP using NaBH$_4$ as reducing agent could be explained using Langmuir-Hinshelwood heterogeneous catalysis model. The superior catalytic activity of Ni-Co LDH nanosheets can be attributed to its unique 2D lamellar structure, which provides enhanced active sites for reactant molecules to adsorb and perform reduction reactions. Quite interestingly, Ni-Co LDH as an electrocatalyst, shows better catalytic activity with remarkable durability for HER. A minimal overpotential of 247 mV at a current density of 10 mA cm$^{-2}$ was displayed by the Ni-Co LDH catalyst in 1 M KOH, This opens many avenues for large scale industrial use of our materials.

**Acknowledgements**

The authors acknowledge the funding received from DST (India) under its MES scheme for the project entitled, "Hierarchically nanostructured energy materials for next generation Na-ion based energy storage technologies and their use in renewable energy systems" (Grant No.: DST/TMD/MES/2k16/77).

# Supplementary Information

**Superior-catalytic performance of Ni-Co Layered double hydroxide nanosheets for the reduction of p-nitrophenol**

Sakshi Kansal[1], Paulomi Singh[1], Sudipta Biswas[2], Ananya Chowdhury[2], Debabrata Mandal[3], Surbhi Priya[1], Trilok Singh[1], Amreesh Chandra[1,2,3*]

[1]*School of Energy Science and Engineering,* [2]*Department of Physics,* [3]*School of Nano Science and Technology*

*Indian Institute of Technology Kharagpur, Kharagpur – 721302, India*

Email: *achandra@phy.iitkgp.ac.in

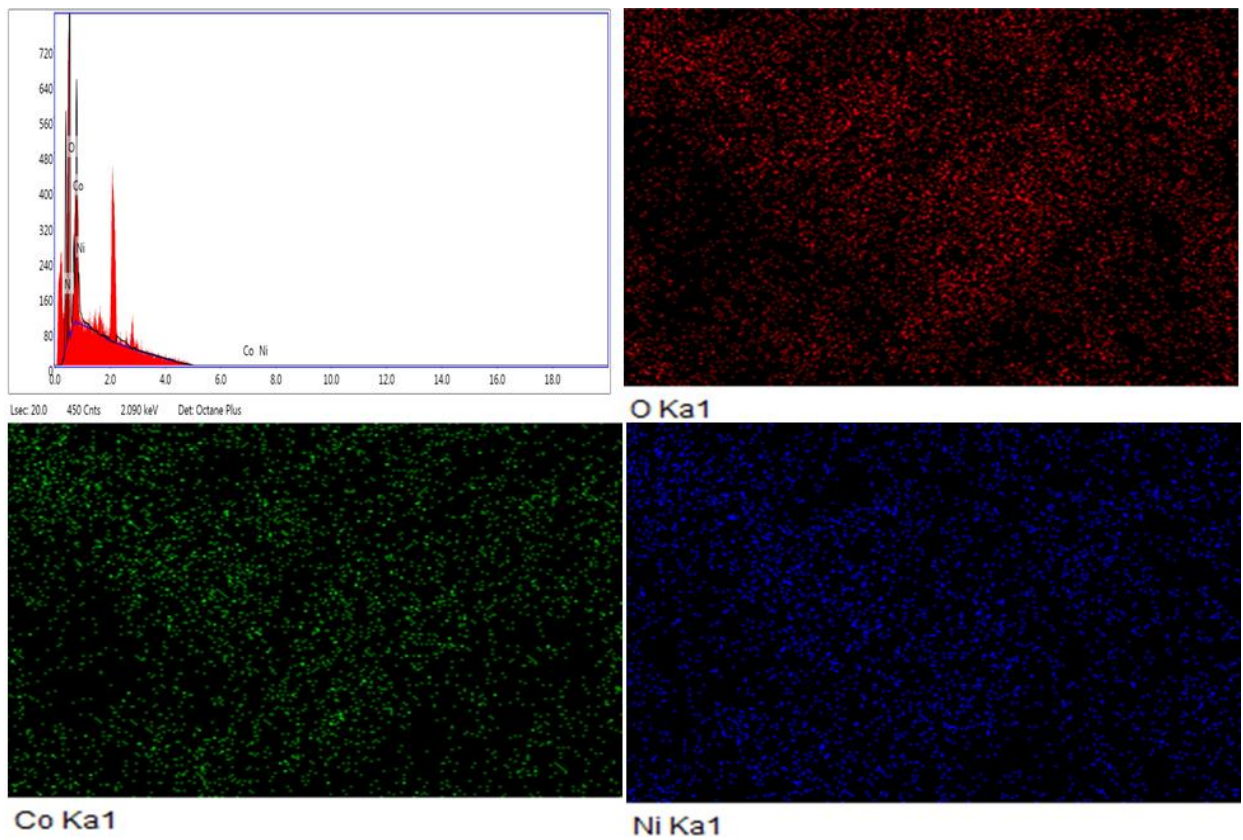

Figure S1 EDX analysis of Ni-Co LDH nanosheets

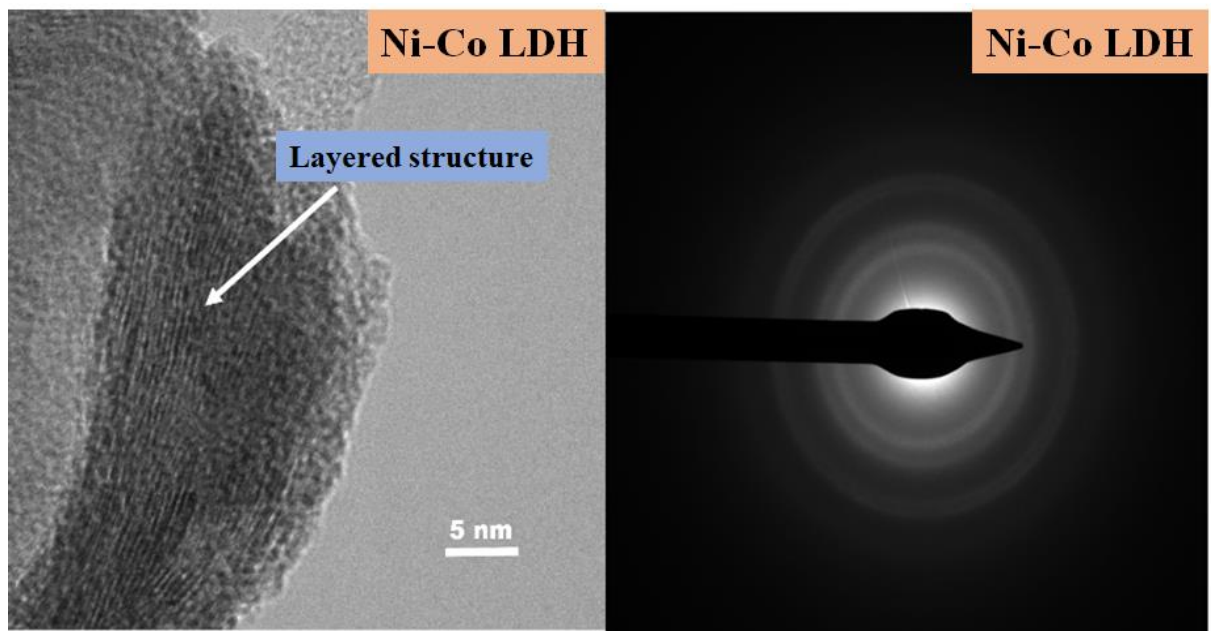

Figure S2 (a) TEM image, (b) SAED pattern of Ni-Co LDH nanosheets

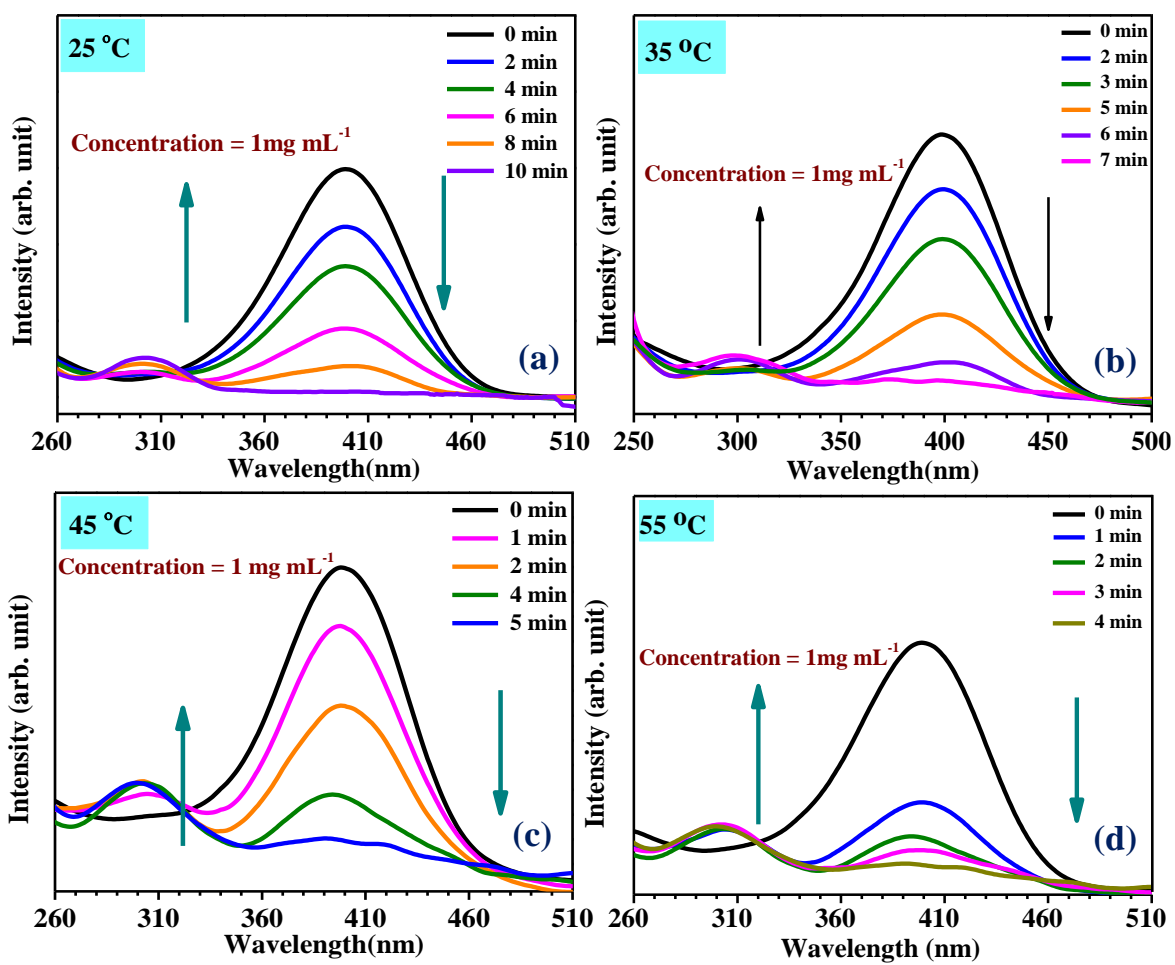

Figure S3 Absorption spectra at (a) 25°C, (b) 35°C, (c) 45°C, and (d) 55°C for p-nitrophenol reduction using Ni-Co LDH nanosheets

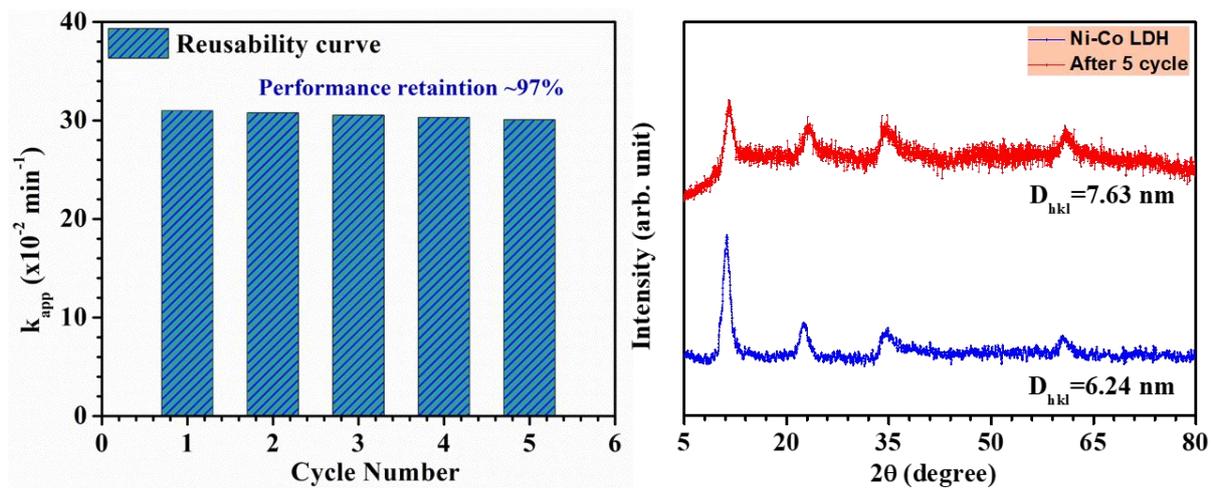

Figure S4 (a) Reusability of Ni-Co LDH nanosheets catalyst6 in successive 5 cycles with 97% retention (b) XRD pattern of Ni-Co LDH before (blue), and after 5 cycles catalysis (red).

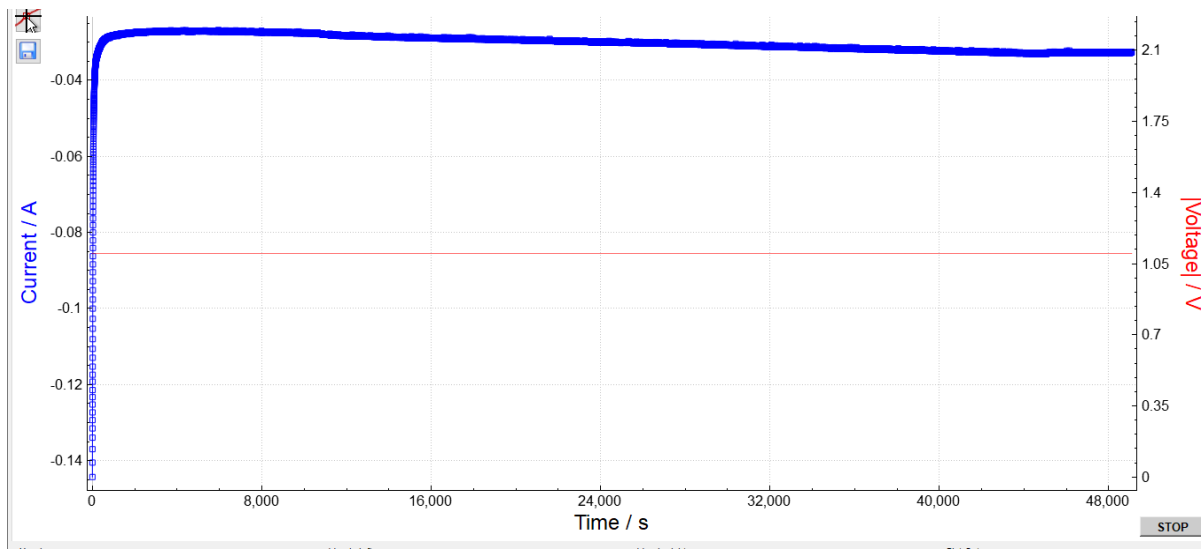

Figure S5 Chronoamperometry done at a constant voltage of -0.15 V for >12 h

**Table S1** Variation of apparent rate constant with different concentration of Ni-Co LDH catalyst

| Concentration of Ni-Co LDH catalyst (mg mL$^{-1}$) | Time taken to complete the reduction (min) | Apparent rate constants($k_{app}$) (min$^{-1}$) |
|---|---|---|
| 0.25 | 27 | 0.03 |
| 0.5 | 18 | 0.14 |
| 1.0 | 10 | 0.31 |

**Table S2** Variation of apparent rate constant and TOF values with same concentration of Ni-Co LDH catalyst at different temperature

| Temperature(°C) | Time interval(min) | Apparent rate constants($k_{app}$) (min$^{-1}$) | TOF×10$^{20}$(molecules g$^{-1}$ min$^{-1}$) |
|---|---|---|---|
| 25 | 10 | 0.31 | 1.20 |
| 35 | 7 | 0.40 | 1.72 |
| 45 | 5 | 0.52 | 2.40 |
| 55 | 4 | 0.65 | 3.01 |

**Table S3** Comparison table of variation of apparent rate constant and TOF values with same concentration of Ni-Co LDH catalyst at different temperature

| Metal-oxide/hydroxide-based catalysts | Electrolyte | Overpotential at 10 mA cm$^{-2}$ (mV) | Tafel slope (mV dec$^{-1}$) | References |
|---|---|---|---|---|
| Co(OH)$_2$@Ni-CoLDH-60 | 1 M KOH | 445 | 307 | 1 |
| EG/Co$_{0.85}$Se/NiFe-LDH | 1 M KOH | 264 | 160 | 2 |
| CoFe LDH-C | 1 M KOH | 415 | 116 | 3 |
| Ni$_{1-x}$Fe$_x$LDH | 1 M KOH | 242 | 110 | 4 |
| Ni$_2$Co-B-P | 1 M KOH | 160 | 98 | 5 |
| NiFe-LDH/FeCoS$_2$/CFC | 1 M KOH | 308 | 157 | 6 |
| **Ni-Co LDH nanosheets** | 1 M KOH | 247 | 230 | This work |